# Presentation of a Recommender System with Ensemble Learning and Graph Embedding: A Case on MovieLens

Saman Forouzandeh[1], Kamal Berahmand[2], Mehrdad Rostami[3]

Department of Computer Engineering University of Applied Science and Technology, Center of Tehran Municipality ICT org.Tehran, Iran[1]
Department of Science and Engineering, Queensland University of Technology, Brisbane, Australia[2]
Department of Computer Engineering, University of Kurdistan, Sanandaj, Iran[3]
Saman.forouzandeh@gmail.com[1], kamal.berahmand@hdr.qut.edu.au [2], m.rostami@eng.uok.ac.ir[3]

**Abstract:**
Information technology has spread widely, and extraordinarily large amounts of data have been made accessible to users, which has made it challenging to select data that are in accordance with user needs. For the resolution of the above issue, recommender systems have emerged, which much help users go through the process of decision-making and selecting relevant data. A recommender system predicts users' behavior to be capable of detecting their interests and needs, and it often uses the classification technique for this purpose. It may not be sufficiently accurate to employ individual classification, where not all cases can be examined, which makes the method inappropriate to specific problems. In this research, group classification and the ensemble learning technique were used for increasing prediction accuracy in recommender systems. Another issue that is raised here concerns user analysis. Given the large size of the data and a large number of users, the process of user needs analysis and prediction (using a graph in most cases, representing the relations between users and their selected items) is complicated and cumbersome in recommender systems. Graph embedding was also proposed for resolution of this issue, where all or part of user behavior can be simulated through the generation of several vectors, resolving the problem of user behavior analysis to a large extent while maintaining high efficiency. In this research, individuals most similar to the target user were classified using ensemble learning, fuzzy rules, and the decision tree, and relevant recommendations were then made to each user with a heterogeneous knowledge graph and embedding vectors. This study was performed on the MovieLens datasets, and the obtained results indicated the high efficiency of the presented method.

## 1- Introduction:

As the Internet is used more widely, users are confronted with an issue known as information overflow, in the sense that there is an excessive amount of information, in such a way that users are lost in the information, and cannot use the data completely to their advantage. Since recommender systems (RSs) have emerged, and information has been customized for users, the above problem has been resolved to a large extent, and users can identify their requirements and preferences using those systems, and achieve their purposes and priorities more rapidly by customizing services [1-4]. Recommender systems are of three broad types including Collaborative Filtering (CF), Content-Based Filtering (CBF), and hybrid systems. CF techniques are popular methods and can be used in most online purchasing websites [5]. A CF system retrieves information for a user assuming that the user likes an item that other users have liked in the past. CF retrieves information for the target user using similarities between items (item-based) or between users (user-based) and makes recommendations according to the obtained results [6]. Recommendations are made in CF based on analyses made of the feedbacks received from users, which can include actions such as ratings, reviews, clicks, and purchases. CBT, however, depends on item metadata for the generation of recommendations [7]. As stated, a CF system needs to analyze user behavior for that purpose to be able to find similar behavior in other users and make recommendations on that basis (based on similarity in user profiles or item selection). The issue raised here concerns how customer behavior is predicted by recommender systems, which often make predictions according to individual classifications, while the generation of an ensemble classification can increase prediction accuracy, and the variety in the recommendations made to users can be increased through a combination of different classes [8]. For this purpose, this research uses ensemble learning, a machine learning method that uses a combination of different models as a single model to improve the results. A combination method is an integral part of an ensemble classifier. The employed combination methods make combination predictions by using the predictions made by the individual classifications and combining them. Predictions based on combinations of classifiers exhibit greater prediction accuracy, which facilitates the decision-making process[9]. An ensemble classifier uses a group decision-making process to make predictions using a set of basic classifications.

It has become common to use various graphs for the representation of the relations between users and their analysis, causing subtle information to be discovered, which can often not be observed and employed without graph analyses. Moreover, their utilization can be very beneficial and crucial. The applications include analysis of various social networks, node classification [10], node clustering [11], node retrieval/recommendation [12], link prediction [13], detection of different people in social networks [14], and recommendation of various items to users. Although it is necessary to analyze various graphs, it requires complex computations and large spaces, because a real graph may contain thousands or millions of nodes on the intended scale [15-17]. An effective solution to the problem involves the use of graph embedding, which resolves the graph analysis issue, and is highly efficient as well. Graph embedding turns a graph into a low-dimensional space and preserves the information on it at the same time[14]. There are different graph types, such as homogeneous graphs, heterogeneous graphs, and attribute graphs, which output a low-dimensional vector, representing the entire graph or part of it [18]. Combined with machine learning algorithms, the learned embedding can be applied to various significate tasks, such as expert finding, relationship prediction, node classification, community identification, etc. Recent years have witnessed the success of the network embedding, such as DeepWalk, LINE, node2vec and other deep neural network-based methods [19, 20]. In this research, graph embedding is used, and parts of the graph are examined that are most similar in behavior and profile to the target user based on the results obtained from ensemble learning. On the basis of the data under study, the graph is traversed, a number of vectors are generated, and recommendations are made according to the obtained output.

Therefore, the method presented in this research is implemented in two steps on the MovieLens datasets. First, different classes are created using group classification and the ensemble learning technique based on the target user, and the similarity (both in profile and in behavior) between the created classes and the target user is examined according to different parameters. Three classes are finally selected (using the output obtained from fuzzy rules and the decision tree) that are most similar to the target user in terms of the defined measure. Then, the target user's behavior toward these three classes of users forms a heterogeneous knowledge graph, and the relations between different nodes on the graph are defined based on several fuzzy rules. Several vectors are generated based on the above relations, and knowledge extraction and system learning are carried out using the node2vec embedding technique. Finally, movies are recommended to the user (through the creation of a User Triple Vector (UTV)) that are among the top-*k* movies and have scored higher than all the others. The results obtained from the output of this research suggest that recommendations are made using this method more accurately than in similar methods (given the results obtained from the assessment). The innovations made here are as follows:

- use of group classification and ensemble learning, which is more accurate than individual classification [8]
- formation of a heterogeneous knowledge graph according to users that are most similar in behavior and profile
- Recommendation to users based on the relations defined between them through the generation of a user triple vector.

The paper is organized as follows: In Section 2, we present a review of the literature on Ensemble Learning and Embedding Techniques on recommendation systems, Section 3 presents the Methodology, we describe the Recommendation Model with Graph Embedding in Section 4, and Section 5 contains the experimental experience.

## 2- Related Works:

The review of the previous works is composed of two sections based on the method used in this research. Works relevant to ensemble learning are reviewed in the first section, and embedding techniques in recommender systems are investigated in the second section.

## 2-1- Ensemble Learning:

Ensemble methods constitute one of the most promising research directions [21]. An ensemble is also known as a multiplex classifier or committee consisting of several separate classifiers that can predict new items by combining the inputs. The application of ensemble methods has demonstrated that prediction accuracy can thus be improved, and complex problems can be solved through division into many simpler sub-problems. The main motivation for using classifier ensembles is the no free lunch theorem formulated by Wolpert [22]. Based on the above theory, there is no classifier available that can be employed individually to solve all problems and algorithms, since each of them involves its domain and topics. We usually have access to one group of classifiers to be able to solve a specific problem. Turner [21] demonstrated that the average output of an infinite number of independent classifiers might lead to the same solution as that of the optimal Bayes classifier [23]. Ho [24] emphasized that a hybrid decision function should obtain useful performance from each decision function. Specifically, they considered several methods according to different levels of decision-making, such as the Borda count. The important points to be noted upon the generation of a classification ensemble include the following [25].

- Proposing interconnections among individual classifiers in the ensemble.
- Selecting a pool of diverse and complementary individual classifiers for the ensemble.
- Proposing a combination rule, responsible for the final decision of the ensemble, which should exploit the strengths of the component classifiers.

The main motivation for using classifier ensembles is the no free lunch theorem formulated by Wolpert [22]. The overall diagram of a classifier ensemble is shown in Figure 1.

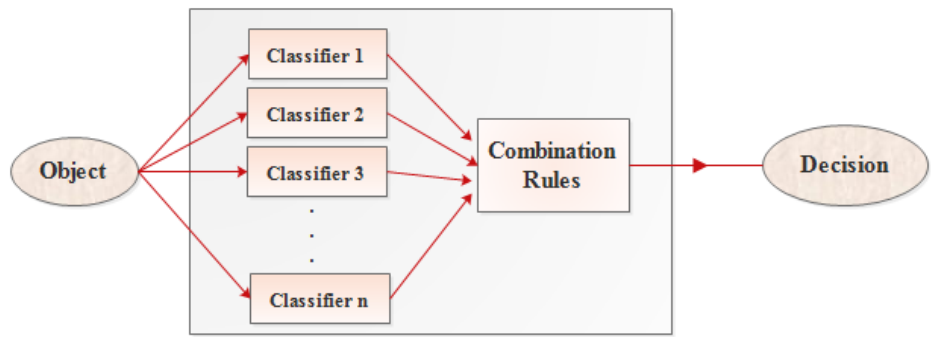

**Fig. 1. Diagram of the classifier ensemble**

It is a key factor in an ensemble classification to select the classifiers properly. An ideal ensemble includes individual classifiers that complement each other, each involving variety and high accuracy [26]. It has been agreed in general that a good ensemble contains individual classifications with high accuracy, where the variety in the examination of different features is effective on the prediction accuracy of the ensemble [27]. Finally, the selection of each of the classifiers and their combination with each other causes the results obtained from the ensemble to be positive and acceptable as compared to those in the individual mode. In [28], a multi-criteria tourist recommendation system based on CF algorithms was designed using cluster ensembles and machine learning techniques for prediction. Moreover, the ANFIS[1] and SVR[2] models were used for the prediction of user behavior. The presented method was assessed over the TripAdvisor dataset, and the obtained results suggested its higher efficiency than that of the individual CF algorithms. In [29], a recommender server was designed in IoT, where the recommender system was trained based on ensemble learning. For this purpose, four classes were generated, one of which included the user profile, where the relevant fields were combined with three other classes, and new states were thus made, and prediction accuracy was obtained for each of the provided combinations. In [30], a drug recommender system was designed using an ensemble model, where the side effects of each drug were predicted by the RS, and it was demonstrated that use of an ensemble increases the performance of an RS in the recommendations made to users and the accuracy of the predictions. The paper [31] proposed an ensembling approach to unify different types of feedback from users when consuming content to provide better recommendations. The advantage is that more information about the interests of the user can be obtained when analyzing multimodal interactions. The authors of [32] demonstrated that a combination of simple techniques could be more effective in the generation of a complex algorithm that functions independently. The authors of [32] applied a systematic framework of ensemble methods to CF recommender systems. They used automatic methods for the creation of a series of collaborative filtering methods based on an ensemble-based algorithm and demonstrated the greater effectiveness of the method than that of CF-based ones. [33] Addressed a hybrid book recommender system using linked open data, combining different strategies for making recommendations. The authors' approach involved the use of individual-based instructions and recommendations to users according to ratings made by other users. The algorithm employed in the research made recommendations to users combining the ensemble method and ranking aggregation. The authors demonstrated that their method causes great diversity in the recommendations made to users.

---
[1] Adaptive Neuro-Fuzzy Inference Systems
[2] Support Vector Regression

## 2-2- Embedding Techniques in Recommender Systems:

The use of various network embedding techniques has gained ground in recent years [34-37] and has become versatile in the resolution of problems concerning graph mining, such as link prediction, vertex classification, and community detection, with a large number of researchers using different network embedding methods. The use of various embedding techniques in different areas of recommender systems has also grown considerably. A recommender system based on friendships, for instance, can predict user interests and preferences [38], and the theory of belief functions can estimate the effects of users in a social network [39]. There can be a very large repository of various wise decisions for jobs such as node classification, link prediction, and information dissemination through the extraction of valuable information on a social network [40]. The most common word and text representations rely on sparse high-dimensional vectors based on one-hot and bags-of-words [7] models, respectively. Nevertheless, recent neural embedding methods have provided dramatic advances in several natural language processing tasks by learning a fixed-length dense vector representation of words [41-43], and texts [44]. GloVe is another successful model for learning word embedding based on global matrix factorization and local context window methods [43]. On the other hand, Le and Mikolov also introduced two models for learning representations of any piece of text [44]. Recently, the word2vec skip-gram model [41, 42] has been used to build item representations in collaborative filtering scenarios [45-47]. Grbovic et al. proposed prod2vec, a model that learns item embedding for product recommendation [45]. Embedding vectors have also been introduced for matrix factorization methods, the commonest involving model-based recommenders [48]. Guardia-Sebaoun et al. proposed to embed temporal information into user and item representations to build time-aware recommender systems [49]. They modeled user actions as ordered sequences of items. They used these sequences to learn item embedding, and they represented each user as a series of movements in the item space. Then, they applied a matrix factorization algorithm to these time-aware representations [7]. In [20], a new method of network embedding known as rank2vec was introduced. The main idea of Rank2vec is to learn the representation of each node by selecting a set of representative node sequences.

In paper [7], they presented prefs2vec, a neural embedding model for collaborative filtering inspired in word2vec continuous bag-of-words (CBOW) model. This model exploits collaborative filtering preferences to compute user and item embedding. Paper [50] proposes a novel RS that is built upon the semantics-based items' content embedding model, enriched with contextual features extracted through Convolutional Neural Network (CNN). Non-negative Matrix Factorization (NMF), supplied with improvised embedding, is used as CF technique [50]. In paper [51], authors apply the knowledge graph techniques to the financial news recommendation task and propose a model for incremental updating of embedding that focuses on computation time and performance. Based on the knowledge graph created to represent users, news, companies, and other entities, nodes' relatedness scores are obtained with node2vec following a feature learning approach based on neural language models. Research [52] proposed an extension of Bordes's work, enabling entities to have different representations in the context of different relations by projecting entities on a hyperplane identified by the normal vector. Lin et al. described a model that enabled entities and relations to be embedded in a vector space with different dimensions through a projection matrix associated with any relation [53]. Perozzi et al. proposed Deep Walk, a method simulating random walks on a graph and generating sequences that were successively processed by a neural language model [36]. Grover et al. introduced node2vec, a more flexible and complex improvement of Deep Walk [35]. In paper [54], researchers propose a method of hashtag recommendation that computes tweet embedding using word2vec features and for representing each tweet by its weighted averaging value, then combining these features with DBSCAN clustering algorithm.

## 3- Methodology:

The procedure of implementation of the present research includes two major steps. In the first step, the MovieLens dataset is studied, and the required fields on the dataset are specified. Some classes are generated using the ensemble learning technique, and the accuracy of each of the recommendations obtained in each class is then computed and assessed using the Precision measure so that classes with the greatest similarity to the target user in terms of profile and behavior can be detected through a combination of the created classes. For this purpose, different combinations of classes are examined using ensemble learning, fuzzy rules, and decision trees, and classes are finally created through the above combinations that exhibit the highest accuracy in the recommendations made to users. In the second step, a heterogeneous knowledge graph is generated after the intended classes are detected to obtain the relations between different nodes through the definition of various embedding vectors. A user triple vector is then formed for each user, the outputs obtained from which are used for making recommendations to users. The proposed method is implemented and adapted on the MovieLens 1M and MovieLens 10M datasets, and the vectors are combined based on the parameters defined there. Then, the implemented method is compared to other recommender system algorithms in the experimental assessment section. Table 1 shows a summary of the information on the datasets under investigation here.

**Table 1. Statistics on the datasets used in the experiments**

| Datasets | Users | Items | Ratings |
|---|---|---|---|
| MovieLens-10M | 71,567 | 10,681 | 10,000,054 |
| MovieLens-1M | 6,040 | 3,952 | 1,000,209 |

## 3-1- Experimental Settings:

In this subsection, each of the classes generated in ensemble learning and the method considered in this research is described. On that basis, the implementation procedure in the first step given the MovieLens dataset is shown in Figure 2.

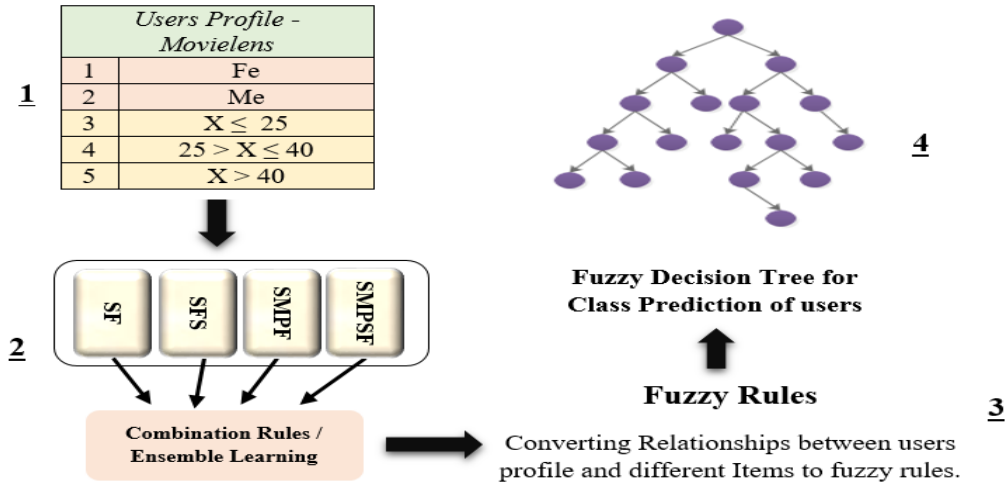
**Fig. 2. Diagram of the ensemble learning**

As observed, the research is conducted in four phases in the first step, where each user profile is analyzed on MovieLens, and each profile, in turn, consists of five fields (the two parameters Gender and Age). Each user profile is given to the system as input, and each field is then combined with various classes created in the second step in Figure 2. There are four different classes in the second step, as detailed below. The four classes are combined with the profile fields concerning each user, and the output obtained from the combination is obtained as fuzzy rules for each user in the third step. In the fourth step, the fuzzy decision tree for the extracted rules is drawn, and the accuracy of each of the created classes in the obtained decision tree is specified. Finally, three classes with the highest accuracy (the highest-profile and behavior similarity) are selected, and various recommendations are made to users with the output obtained from this step and graph embedding. The generated classes are defined below.

### 3-2- Definitions and Preliminaries:
In this subsection, it is described how the four classes considered in Figure 2 are created given the method presented in the first step of the research. Thus, the created classes are as follows.

### 3-2-1- Similarity in User Profiles (SUP):
This class is generated based on user profiles. A user profile is composed of the two fields Gender and Age in this research, which constitutes a total of five states (given age groups A, B, and C), as follows:
- Gender (Male, Female)
- Age (X ≤ 25, 25 < X ≤ 40, X > 40).

In Formula 1, therefore,

$$\text{Pro2Vec} = \sum_{i=2}^{i=2}[\text{Max (Male or Female)}. \text{Max} (X \leq 25 \text{ or } 25 > X \leq 40 \text{ or } X > 40)$$

⟹ Pro2Vec = Male & (X ≤ 25) - Male & (25 > X ≤ 40) – Male & ( X > 40) - Female & ( X ≤ 25 ) - Male & (25 > X ≤ 40) – Male & (X > 40)  (1).

Based on the above formula, different combinations of different user profile fields are generated and combined with the other four classes, introduced below, on which basis the similarity between users (in terms of profile and behavior) can be examined.

### 3-2-2- Similarity in Film Selection (SF):
This class is used to obtain users with the greatest similarity in behavior and item ratings. Thus,

$$\text{SF} = \sum_{i \varepsilon L_{uv}} (r_{ui})^2 (r_{vi})^2 \quad (2).$$

Use of Formula 2 allows us to specify the items rated by both user's $u$ and $v$ to obtain their shared rates. $r_{ui}$ and $r_{vi}$ represent the rates given to item $i$ by users $u$ and $v$. Given that the purpose of this research is to recommend items to users following their interests, it needs to be specified for each user what scores suggest his interest, and what scores suggest his lack of interest in items shared by users. For this purpose, the mean score given by the users in a movie rating is obtained, on which basis Formulae 3 and 4 hold:

$$\text{Interests} = X \geq \sum_{i \in n}^{i=n} \lfloor r_{ui}/n \rfloor \quad (3).$$
$$\text{Lack of interest} = X < \text{Interests} \quad (4).$$

In Equation 3, $r_{ui}$ denotes that item $i$ has been rated by user $u$ and scored $r$, and $n$ indicates the number of items that have been rated. Since the assigned scores are integers, the minimum absolute value of the mean obtained from the user rating is considered. Thus, the scores given by the users to the shared items are compared, which allows us to specify interest or lack of interest in those items. If the score given to a movie is equal to or greater than the user's mean score in the movie rating, the user is interested in the movie, and if it is less than the value obtained for Interest, the user is not interested in the specific item or movie.

### 3-2-3- Similarity in Film Subjects (SFS):

This class contains items (movies) in which users sharing a subject with the target user are interested. The target user has rated a movie in the genre of Action or Drama, say, and the score given to the genre indicates his interest in the movie. In this class, the output involves the behavior of users that have expressed interest in movies sharing a subject (genre) with those in which the target user is interested. Therefore, Formula 5 holds:

$$\text{SFS} = \sum_{i=1}^{i=n} \left( U_{F_{x_i} \cdot F_{y_i}} \ldots n \right) \quad (5).$$

Where $n$ represents the number of users, and there is at least one user, $U$ indicates the target user, and $F_{x_i}$ and $F_{y_i}$ show movies $X$ and $Y$, both relevant to subject $i$.

### 3-2-4- Similarity in Selection of Most Popular Films (SMPF):

This class contains users who have selected the most popular movies. The most popular movies are chosen based on the number of their selections by the users. Therefore, the mean scores given to the movies by the users are calculated, and popular movies are identified according to their obtained scores on Formula 6, allowing the items (movies) to be rated in terms of popularity:

$$P_F = \sum_{i=Rating}^{i=Items} \lfloor r_{ui} / n \rfloor \quad (6).$$

Thus, movies are regarded as popular that have most frequently been given the assumed minimum score by the users based on the above formula (The range of selected items is the number of movies rated by the users and the total number of ratings made by them). For further comparison of the data concerning the intended user's behavior and those about the other users' behavior in this research, movies rated second to fifth in terms of popularity are considered as well as the most popular one. Hence, the similarity between the behaviors of users who have selected at least one of the five most popular movies is taken into account. Therefore, Formula 7 holds:

$$\text{SMPF} = \sum_{i=1}^{i=5} \left(U_{F_1}\right) \text{ or } \left(U_{F_2}\right) \text{ or } \left(U_{F_3}\right) \text{ or } \left(U_{F_4}\right) \text{ or } \left(U_{F_5}\right) \quad (7)$$

Where $U$ represents the users who have selected one of the most popular movies $F_1$, $F_2$, $F_3$, $F_4$, and $F_5$. The operator OR is used between the five popular movies so that all the users who have selected at least one of them can be chosen, as the users are filtered.

### 3-2-5- Similarity in Selection of Films with Most Popular Subjects (SMPSF):

Users are placed in this class who have selected items belonging to the most popular subjects or genres, which in turn includes movies with subjects or genres selected most frequently by the users. Therefore, this class contains users with shared behavior in the selection of movies concerning the most popular subjects, as stated in Formula 8:

$$\text{MPSItems2Vec} = \sum_{i=1}^{i=5} \left(U_{F_{S_1}}\right) \text{or } (U_{F_{S_2}}) \text{ or } (U_{F_{S_3}}) \text{ or } (U_{F_{S_4}}) \text{or } (U_{F_{S_5}}) \quad (8)$$

Where the movie *F* concerns one of the five subjects $S_1$, $S_2$, $S_3$, $S_4$, and $S_5$, which are the most popular and have been selected by the users *U*.

### 3-3- Ensemble Learning:

Ensemble learning is a machine learning method where a combination of identical models is provided as a single model for improvement of the final result. An ensemble model is composed of *k* learners (of the classifier or predictor type) as $k_1, k_2, ...., k_n$ for the generation of an improved hybrid model as compared to individual classifiers [55]. The recommender system presented in this research creates classes of users based on an ensemble using individual pieces of learning and then compares the behavior of the users in each class to that of the target user. Once each classification scheme is specified, each of the classifiers is learned in the following step. For this aim, the data need to be labeled, on which basis a number of classes are created, and the accuracy of each is obtained. To that aim, the user profile fields are combined with four other classes in the second step in Figure 2, and a new class is generated each time for filtering the users, obtaining the accuracy of each created combination. On that basis, the classifiers can create twelve combinations given the user profiles and four created classes, as follows.

1. This class contains users who are similar to the target user in profile and have selected shared movies as interest.
$$U_1 = \sum U_{SUP} \,\&\&\, U_{SF} \qquad (9).$$

2. This class contains users who are similar to the target user in profile in the fields Gender and Age and have selected movies as the interest that share subjects with those selected by the target user as interest.
$$U_2 = \sum U_{SUP} \,\&\&\, U_{SFS} \qquad (10).$$

3. This class contains users who are similar to the target user in profile and have selected movies as the interest that is regarded as popular among the movies.
$$U_3 = \sum U_{SUP} \,\&\&\, U_{SMPF} \qquad (11).$$

4. This class contains users who are similar to the target user in profile and have selected movies as the interest that concern subjects regarded as popular among the movies.
$$U_4 = \sum U_{SUP} \,\&\&\, U_{SMPSF} \qquad (12).$$

5. Users with shared profiles who have selected shared movies concerning shared subjects as interest are grouped in this class.
$$U_5 = \sum U_{SUP} \,\&\&\, U_{SF \,\&\, SFS} \qquad (13).$$

6. Users with profile similarity to the target user who have selected the most popular movies concerning the most popular subjects are grouped in this class.
$$U_6 = \sum U_{SUP} \,\&\&\, U_{SMPF \,\&\, SMPSF} \qquad (14).$$

7. Users with profile similarity to the target user who have selected the most popular movies with a shared subject are grouped in this class.
$$U_7 = \sum U_{SUP} \,\&\&\, U_{SMPF \,\&\, SFS} \qquad (15).$$

8. This class contains users who are similar to the target user in profile, and have selected movies concerning the most popular subjects, and liked movies that share subjects with those selected by the target user.
$$U_8 = \sum U_{SUP} \,\&\&\, U_{SMPSF \,\&\, SFS} \qquad (16).$$

9. Users with profile similarity to the target user who have selected the most popular movies, and are interested in shared movies are grouped in this class.
$$U_9 = \sum U_{SUP} \,\&\&\, U_{SMPF \,\&\, SF} \qquad (17).$$

10. This class contains users who are similar to the target user in profile, have selected movies concerning the most popular subjects, and are interested in shared movies.
$$U_{10} = \sum U_{SUP} \,\&\&\, U_{SMPSF \,\&\, SF} \qquad (18).$$

11. This class contains users who are similar to the target user in profile and have selected the most popular movie, ones concerning the most popular subjects, and movies sharing subjects with those selected by the target user.
$$U_{11} = \sum U_{SUP} \,\&\&\, U_{SMPF \,\&\, SMPSF \,\&\, SFS} \qquad (19).$$

12. Users with profile similarity to the target user who have selected the most popular movie, ones concerning the most popular subjects, and shared movies are grouped in this class.

$$U_{12} = \sum U_{SUP} \,\&\&\, U_{SMPF\,\&\,SMPSF\,\&\,SF} \quad (20).$$

Therefore, the final film ensemble learning functions according to Formula 21:

$$FEL = \sum_{i=1}^{i=2} U_{Max\,(Gender.\,Age)} \text{ and } \sum_{i=1}^{i=4}[((U_{SF} \cdot U_{SFS} \cdot U_{SMPF} \cdot U_{SMPSF})] \quad (21).$$

Based on the above formula, the final FEL is composed of two general parts. The first part contains the user profile, the fields of which are combined with four other classes to generate the final output. $\sum$ User Ranges from 1 to 2 in this formula, since data may be filtered based on a minimum of one and a maximum of two user profile fields given user status. This is true also of the second part of the formula, where a minimum of one and a maximum of four classes are combined with the user profile. The Precision measure is used below for investigation of the accuracy of the generated fuzzy rules, and user behavior similarity in the selection of different items is thus assessed. That is, the higher the value of Precision, the greater the behavior similarity between the users grouped in this class. On that basis, Formula 18 holds [56, 57]:

$$Precision = \frac{TP}{TP+FP} \quad (22).$$

Where TP is the number of items shared by the users, indicating the value of their shared behavior (in terms of the number of items, and FP is the number of items involving nothing shared by the users. That is, if the users are grouped in the same class concerning the created combinations, and have rated a total of 100 items, 55 of which are shared by them, the following equation holds:

$$Precision = \frac{55}{55+45} = 55\%.$$

An example is addressed in this subsection for further investigation of the issue, where a user with the profile (Gender = Female, 25 < Age ≤ 40) is examined on the MovieLens dataset. After the system is trained, and the user profile is combined with four defined classes, the generated fuzzy rules along with the value of Precision for each rule are as shown in Table 2.

**Table 2. Fuzzy rules, [User 1: (Female, 25 < X ≤ 40)]**

| |
|---|
| 1: IF [Gender = Female] OR [Gender = Male] AND [SF = True] Then Precision = 12% |
| 2: IF [Gender = Female] OR [Gender = Male] AND [SFS = True] Then Precision = 19% |
| 3: IF [Gender = Female] OR [Gender = Male] AND [SMPF = True] Then Precision = 17% |
| 4: IF [Gender = Female] OR [Gender = Male] AND [SMPSF = True] Then Precision = 22% |
| 5: IF [Gender = Female] OR [Gender = Male] AND [SF = True] AND [SFS = True] Then Precision = 26% |
| 6: IF [Gender = Female] OR [Gender = Male] AND [SF = True] AND [SMPF = True] Then Precision = 29% |
| 7: IF [Gender = Female] OR [Gender = Male] AND [SF = True] AND [SMPSF = True] Then Precision = 19% |
| 8: IF [Gender = Female] OR [Gender = Male] AND [SFS = True] AND [SMPF = True] Then Precision = 30% |
| 9: IF [Gender = Female] OR [Gender = Male] AND [SFS = True] AND [SMPSF = True] Then Precision = 29% |
| 10: IF [Gender = Female] OR [Gender = Male] AND [SMPF = True] AND [SMPSF = True] Then Precision = 35% |
| 11: IF [Gender = Female] OR [Gender = Male] AND [SF = True] AND [SFS = True] AND [SMPF = True] Then Precision = 47% |
| 12: IF [Gender = Female] OR [Gender = Male] AND [SF = True] AND [SFS = True] AND [SMPSF = True] Then Precision = 50% |
| 13: IF [Gender = Female] OR [Gender = Male] AND [SF = True] AND [SMPF = True] AND [SMPSF = True] Then Precision = 58% |
| 14: IF [Gender = Female] OR [Gender = Male] AND [SFS = True] AND [SMPF = True] AND [SMPSF = True] Then Precision = 65% |
| 15: IF [Gender = Female] OR [Gender = Male] AND [SF = True] AND [SFS = True] AND [SMPF = True] AND [SMPSF = True] Then Precision = 73% |
| 16: IF [Gender = Female] OR [Gender = Male] AND [Age = (X ≤ 25)] OR [Age = (25 > X ≤ 40)] OR [Age = (X > 40)] AND [SF = True] Then Precision = 19% |

17: IF [Gender = Female] OR [Gender = Male] AND [Age = (X ≤ 25)] OR [Age = (25 > X ≤ 40)] OR [Age = (X > 40)] AND [SFS = True] Then Precision = 18%

18: IF [Gender = Female] OR [Gender = Male] AND [Age = (X ≤ 25)] OR [Age = (25 > X ≤ 40)] OR [Age = (X > 40)] AND [SMPF = True] Then Precision = 20%

19: IF [Gender = Female] OR [Gender = Male] AND [Age = (X ≤ 25)] OR [Age = (25 > X ≤ 40)] OR [Age = (X > 40)] AND [SMPSF = True] Then Precision = 17%

20: IF [Gender = Female] OR [Gender = Male] AND [Age = (X ≤ 25)] OR [Age = (25 > X ≤ 40)] OR [Age = (X > 40)] AND [SF = True] AND [SFS = True] Then Precision = 31%

21: IF [Gender = Female] OR [Gender = Male] AND [Age = (X ≤ 25)] OR [Age = (25 > X ≤ 40)] OR [Age = (X > 40)] AND [SF = True] AND [SMPF = True] Then Precision = 71%

22: IF [Gender = Female] OR [Gender = Male] AND [Age = (X ≤ 25)] OR [Age = (25 > X ≤ 40)] OR [Age = (X > 40)] AND [SF = True] AND [SMPSF = True] Then Precision = 49%

23: IF [Gender = Female] OR [Gender = Male] AND [Age = (X ≤ 25)] OR [Age = (25 > X ≤ 40)] OR [Age = (X > 40)] AND [SFS = True] AND [SMPF = True] Then Precision = 42%

24: IF [Gender = Female] OR [Gender = Male] AND [Age = (X ≤ 25)] OR [Age = (25 > X ≤ 40)] OR [Age = (X > 40)] AND [SFS = True] AND [SMPSF = True] Then Precision = 44%

25: IF [Gender = Female] OR [Gender = Male] AND [Age = (X ≤ 25)] OR [Age = (25 > X ≤ 40)] OR [Age = (X > 40)] AND [SMPF = True] AND [SMPSF = True] Then Precision = 51%

26: IF [Gender = Female] OR [Gender = Male] AND [Age = (X ≤ 25)] OR [Age = (25 > X ≤ 40)] OR [Age = (X > 40)] AND [SF = True] AND [SFS = True] AND [SMPF = True] Then Precision = 66%

27: IF [Gender = Female] OR [Gender = Male] AND [Age = (X ≤ 25)] OR [Age = (25 > X ≤ 40)] OR [Age = (X > 40)] AND [SF = True] AND [SFS = True] AND [SMPSF = True] Then Precision = 57%

28: IF [Gender = Female] OR [Gender = Male] AND [Age = (X ≤ 25)] OR [Age = (25 > X ≤ 40)] OR [Age = (X > 40)] AND [SF = True] AND [SMPF = True] AND [SMPSF = True] Then Precision = 51%

29: IF [Gender = Female] OR [Gender = Male] AND [Age = (X ≤ 25)] OR [Age = (25 > X ≤ 40)] OR [Age = (X > 40)] AND [SFS = True] AND [SMPF = True] AND [SMPSF = True] Then Precision = 79%

30: IF [Gender = Female] OR [Gender = Male] AND [Age = (X ≤ 25)] OR [Age = (25 > X ≤ 40)] OR [Age = (X > 40)] AND [SF = True] AND [SFS = True] AND [SMPF = True] AND [SMPSF = True] Then Precision = 69%

As observed, each of the user profile fields is combined with four other classes, and the accuracy value of the class is specified for each combination. Based on Table 1, the combinations shown in lines 15, 21, and 29 are of higher accuracy than the other combinations, with accuracy values of 71%, 73%, and 79%, respectively. The combination in line 29 of Table 1, for instance, involves users who are female, whose ages vary between 25 and 40 years, and who have selected SFS, SMPF, and SMPSF. The decision tree concerning Table 1 is shown in Figure 3.

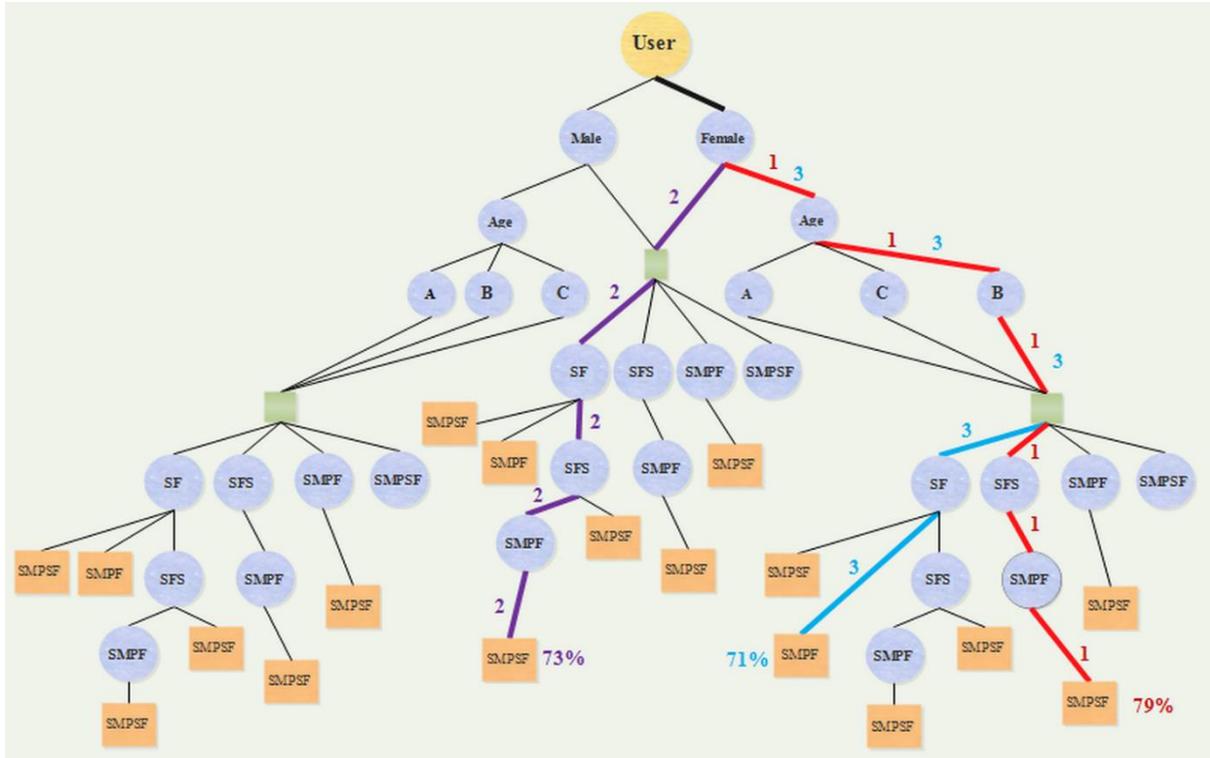

**Fig. 3. The decision tree is drawn for user classification**

As observed, a branch is specified from the tree root to each of the nodes for each row of Table 1. In this tree, user age is shown as three groups: A = X ≤ 25, B = 25 < X ≤ 40, and C = X > 40. The best path can be observed in row 29 with an accuracy of 79%, as shown on the tree by red edges and labels of 1, the second path (in row 15 of the table with an accuracy of 73%) is shown in violet and labels of 2, and the third path (in row 21 of the table with an accuracy of 71%) is shown in blue and labels of 3.

## 4- Recommendation Model with Graph Embedding:

Various recommendations are made concerning the target user, given the results obtained in the section on the fuzzy rules and decision tree. For making recommendations to the users in this section, the heterogeneous knowledge graph and embedding vectors are used. In many works concerned with the analysis of networks and graphs, the graph nodes are not regarded as identical, and each node is of certain importance. In user behavior analysis, for instance, some nodes are considered more closely, since they are of greater importance to that certain user. It is on this basis that there are different edges between nodes in a heterogeneous network, indicating different node relations. In a bibliographical network, for instance, if the nodes represent authors, the edges generated between the nodes can indicate that the authors work in shared areas. The interaction between various nodes usually contains plenty of information, which can be used in network structural learning [58, 59]. Therefore, the structure of the method used in the second step is as shown in Figure 4.

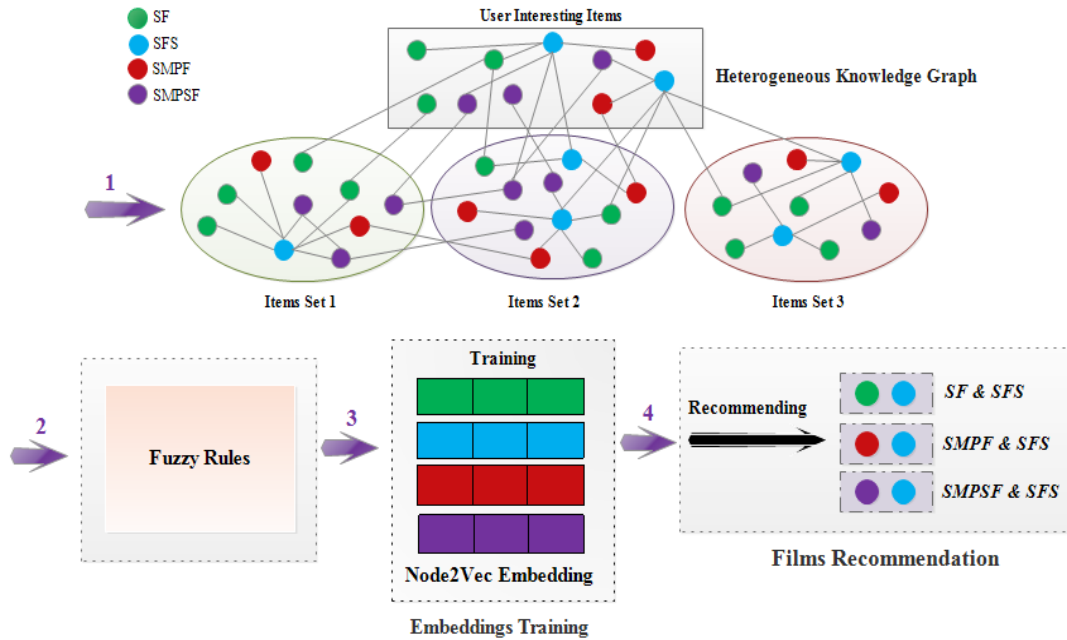

Fig. 4. Conceptual model of making recommendations to users in graph embedding

### 4-1- Node2vec-Based Movie Recommendation:

Node2vec is a generalization of DeepWalk that explores biased random walks. Node2vec introduces two parameters to control the random walk. The return parameter p and the in-out parameter q. The return parameter controls the probability of returning to a node just visited before, and the in-out parameter balances the walk between Breadth-first search (BFS) and Depth-first search (DFS). In node2vec, we learn a mapping of nodes to a low-dimensional of latent space of features that maximizes the likelihood of preserving network neighborhoods of nodes [35]. And it defines a flexible notion of a node's network neighborhood and designs a biased random walk procedure.

The output generated by node2vec exhibits both high accuracy and low time complexity. The algorithm follows two principles:
1. ability to learn representations that embed nodes from the same network community closely together
2. To learn representations where nodes that share similar roles have similar embedding.

The key contribution is in defining a flexible notion of a node's network neighborhood by developing a family of biased random walks, which efficiently explore diverse neighborhoods of a given node. An overview of the Node2vec-based film recommendation method is shown in Fig. 4. This model consists of five parts: graph construction, adjacency matrix, fuzzy rules, training embedding, and film recommendation. For the establishment of relations between the users and movies, a knowledge graph concerning the nodes first needs to be generated, composed of nodes of four types, including SF, SFS, SMPF, and SMPSF, among which various relations are defined, such as relations between movies belonging to a certain genre. Links are established between nodes based on the relations that are there between nodes that represent one type of movie each, and the relations between the users and movies are created indirectly as a result. The intended knowledge graph is generated based on user interest and the output obtained in Section 3-3 for the three classes of users with the greatest similarity to the user, as described in detail below. Node2vec is utilized to generate embedding of the entire graph and make recommendations based on the embedding of users and films according to the cosine similarity between the embedding that are used to calculate the preference order of the candidates.

### 4-2- Heterogeneous knowledge graph:

The heterogeneous knowledge diagram is composed of four parts: the sub-graph of target user interest, sub-graph of first-class users, whose interests are similar to the intended user's with an accuracy of 79%, sub-graph of second-class users, whose interests are similar to the target user's with an accuracy of 73%, sub-graph of third-class users, whose interests are similar to the intended user's with an accuracy of 71% (given Table 2 and Figure 3). Each sub-graph contains the items (movies) that the users in the corresponding class are interested in. Each of the items is divided in turn to four types, including SF, SFS, SMPF, and SMPSF, and the relations between them among different sub-graphs are obtained. The user preferences can be acquired through the established relations, on which basis the relevant recommendations can be made to the user. In Figure 4 and the generated heterogeneous knowledge graph, the green nodes (SF) represent similar items that the users in each group are interested in, and there are neighborhood links between these nodes if they are among the selections made by each of the users. The blue nodes (SFS) indicate subjects and areas that are popular, and there are neighborhood links between these nodes if the green nodes, in which the users are interested, a concern shared areas. For instance, there are links between certain movies that belong to the same genre or subject. More specifically, movies No. 17, 248, and 515 are very popular in each class, and all three belong to the Action genre, so there are links between the three movies. The red nodes (SMPF) indicate the most popular items, and there are relations between these items or movies if they share areas or subjects. That is, there is a neighborhood link between two red nodes sharing a subject (such as two popular nodes that belong to the Drama genre, as they belong to the same genre). The violet nodes (SMPSF) represent items that the users are interested in and belong to popular areas. There are also neighborhood links between violet nodes that belong to a shared area and the relevant area. On that basis, the following definitions hold, where the relations between the nodes on the graph are extracted using fuzzy rules, on which basis relations are then established between nodes that are linked to each other. Based on four defined fuzzy rules, relations are established between the nodes (movies) on different sub-graphs. Therefore, the fuzzy rules defined for the relations between the nodes are as follows:

1. **(SF & SF), Relevant sub-graph.**
If ($U_{SF_i} = S1_{SF_j}$ OR $U_{SF_i} = S2_{SF_j}$ OR $U_{SF_i} = S3_{SF_j}$ OR $S1_{SF_i} = S2_{SF_j}$ OR $S1_{SF_i} = S3_{SF_j}$ OR $S2_{SF_i} = S3_{SF_j}$) Then E= [i][j] = 1

2. **(SF & SFS), Relevant sub-graph.**
If ($U_{SF_i} = S1_{SFS_j}$ OR $U_{SF_i} = S2_{SFS_j}$ OR $U_{SF_i} = S3_{SFS_j}$ OR $S1_{SF_i} = S2_{SFS_j}$ OR $S1_{SF_i} = S3_{SFS_j}$ OR $S2_{SF_i} = S3_{SFS_j}$) Then E= [i][j] = 1

3. **(SMPF & SFS), Relevant sub-graph.**
If ($U_{SMPF_i} = S1_{SFS_j}$ OR $U_{SMPF_i} = S2_{SFS_j}$ OR $U_{SMPF_i} = S3_{SFS_j}$ OR $S1_{SMPF_i} = S2_{SFS_j}$ OR $S1_{SMPF_i} = S3_{SFS_j}$ OR $S2_{SMPF_i} = S3_{SFS_j}$) Then E= [i][j] = 1

4. **(SMPSF & SFS), Relevant sub-graph.**
If ($U_{SMPSF_i} = S1_{SFS_j}$ OR $U_{SMPSF_i} = S2_{SFS_j}$ OR $U_{SMPSF_i} = S3_{SFS_j}$ OR $S1_{SMPSF_i} = S2_{SFS_j}$ OR $S1_{SMPSF_i} = S3_{SFS_j}$ OR $S2_{SMPSF_i} = S3_{SFS_j}$) Then E= [i][j] = 1

Thus, the conditions for the establishment of relations between different nodes are defined based on the above fuzzy rules, and the system is then trained using node2vec embedding.

### 4-3- Embedding Training:
The knowledge graph $G = (V, E)$ is used to learn from nodes that a mapping function allows for feature representations $f: V \rightarrow R^d$ in the form of a matrix with $|V| \times d$ parameters. For every source node u ∈ V, $N_s$ (u) ∈ V is defined as a network neighborhood of node $u$ generated with a neighborhood sampling strategy $S$.
Then, the objective function is optimized, which maximizes the log-probability of observing a network neighborhood $NS$ (u) for node $u$ conditional on its feature representation given by $f$ [7, 59]:

$$\max_f \sum_{u \in V} \log P_r (N_s (u)| f (u)) \qquad (23).$$

In order to make the optimization problem tractable, we make two standard assumptions:

- **Conditional independence.** We factorize the likelihood by assuming that the likelihood of observing a neighborhood node is independent of observing any other neighborhood node given the feature representation of the source [35]:

$$P_r\ (N_s\ (u)|\ f\ (u)) = \prod_{n_i \in N_S(U)} log\ P_r\ (n_i|\ f\ (u)) \quad (24).$$

- **Symmetry in feature space.** A source node and neighborhood node have a symmetric effect over each other in feature space. accordingly, we model the conditional likelihood of every source-neighborhood node pair as a softmax unit parametrized by a dot product of their features:

$$P_r\ (n_i|\ f\ (u)) = \frac{exp(f(n_i).\ f(u))}{\sum_{v \in V} exp(f(v).f(u))} \quad (25).$$

Given conditional independence and symmetry in the feature space, Eq. (26) can be written as follows:

$$\max_f \sum_{u \in V} [\ -\ log\ Z_u + \sum_{n_i \in N_S(U)} f(n_i).\ f(u)] \quad (26).$$

The per-node partition function, $Z_e = \sum_{v \in V} exp\ f\ (U).f\ (V)$, is expensive to compute for large networks and we approximate it using negative sampling [42]. The optimization is performed using stochastic gradient ascent over the parameters defining f, which attempts to maximize the dot product between vectors of the same neighborhood.

### 4-4- Movie Recommendation:

In this part, the node2vec-based Films recommendation (NFR) model is proposed that uses the embedding of movies and users to obtain the relevance between users and different items and to complete the task of film recommendation subsequently. To calculate the preference order of the films, a distance between two nodes is introduced. As Eq. (27) shows, the cosine similarity between the embedding of a pair of nodes is used since the relatedness between them can be easily computed [7].

$$Cos\ (V_1,\ V_2) = \frac{V_1.V_2}{|V_1| \times |V_2|} = \frac{\sum_{j=1}^d (V_{1j} \times V_{2j})}{\sqrt{\sum_{j=1}^d (V_{1j})^2} \times \sqrt{\sum_{j=1}^d (V_{2j})^2}} \quad (27).$$

Above, *d* is the dimension of embedding.

Recommendations are made to the user based on the above four rules, and the other items relevant to and dependent on the user's interests on that basis are detected given the items that the user is interested in, with recommendations made to the user according to the relations defined between different items. For this purpose, a user quadruple vector (UTV) composed of four parts is generated, each of which is detailed. Formula 24 states that items or movies should be detected from among the user's interests that have the same subjects as the items or movies of the user's interest in the other three classes ($u_{SF\ \&\ SFS}$), and the top-k movies are thus selected:

$$S(u_{SF\ \&\ SFS}, n_i\ ) = cos\ (u_{SF\ \&\ SFS}, n_i) \quad (28).$$

Therefore, films n with the top-k highest ranking scores are subsequently selected as the recommendation.

Formula 25 states that if there are other popular items shared by the present and the other three classes among the user's interests, other popular items with areas in common with that in which the target user is interested ($u_{SMPF\ \&\ SFS}$) should be recommended to the user, and the top-k movies are thus selected:

$$S(u_{SMPF\ \&\ SFS}, n_i\ ) = cos\ (u_{SMPF\ \&\ SFS}, n_i) \quad (29).$$

Formula 26 states that if the movies of the user's interest concern areas or subjects that are among the popular subjects in the other three classes, compared with the user's, other movies concerning that popular subject ($u_{SMPSF\ \&\ SFS}$) should be recommended to the user, and the top-k movies are thus selected:

$$S\ (u_{SMPSF\ \&\ SFS}, n_i) = cos\ (u_{SMPSF\ \&\ SFS}, n_i) \quad (30).$$

Where $U$ is the representation of user u, and $SF \& SFS$, $SMPF \& SFS$, $SMPSF \& SFS$ are the films considered a candidate.

$$N_u = \sum_{k=1}^{k} argmax_{n_i \in N} S(u_{[(SF \& SFS) \, or \, (SMPF \& SFS) \, or \, (SMPSF \& SFS)]}, n_i) \quad (31).$$

Eq. (27) is used to choose films to be recommended, where $K$ is the number of films needed, $U$ is the set of target users, $N$ is the set of candidate films, and $N_u$ is the recommendation result for a given user. On that basis, the items concerning each user's interests are identified given Equations 24, 25, and 26, and an item is then selected based on Equation 27 and recommended to the user that is among the top-k movies and has scored higher than other movies or items (given the users' ratings of the movie). As observed, the use of this method makes it possible to examine the items concerning users' interests through an assessment of the behavior of fewer users and to make recommendations to the user on that basis. Another point to be mentioned concerns the amount of change in the vectors embedded for the users, where fewer changes are made than when all the items are examined, and several patterns can be obtained after a while, making it possible for similar users to utilize the available patterns. This greatly improves the time needed for analysis of the items available in various datasets, and recommendations are therefore made to the users more rapidly.

## 5- Experimental Results:

The presented method is evaluated and compared to other methods in this section, containing the following subsections.

## 5-1- Datasets and Experimental Setting:

The proposed method is evaluated using two large datasets. These datasets include MovieLens 1M and MovieLens 10M, which are often used for the recommendation of movies and evaluation of collaborative filtering. In order to demonstrate the performance of the algorithm, we have made the data sparser [60, 61]. MovieLens is a system for a movie recommendation, which is developed by the GroupLens research group at the University of Minnesota. The dataset consists of tuples in the form {user, movie, rating, timestamp}[3], and rating matrix is formed by users on their watched movies, the system predicts the unknown ratings based on observed ratings in the rating matrix, then it recommends users with movies they have never seen and might be interested in. Each rating is an integer between 1 (worst) and 5 (best).

## 5-2- Evaluation Metrics:

For evaluation of the accuracy of the recommendations made to the users, four measurement criteria, known as Precision, Recall, Accuracy (ACC), F-Measure, and RMSE[3], are used. The Precision criterion was used earlier for calculation of the accuracy of each of the generated classes and is used in this subsection for calculation of the accuracy of the recommender system based on a user triple vector. In the formula, TP represents the number of records on which the recommender system has made correct predictions (the number of users whose interests in selection of servers have been predicted correctly), and FP indicates the number of records that have not been selected as TP set members (that have not been predicted correctly by the system, in other words). We make the same discussion about Recall, except that the total number of data is calculated in the prediction there). The following hold, therefore [50]:

$$Precision = \frac{TP}{TP+FP} \quad (22).$$
$$Recall = \frac{TP}{TP+FN} \quad (32).$$
$$Accuracy = \frac{TP+TN}{TP+FP+TN+FN} \quad (33).$$

Where,
- TP represents true positive (i.e. Item is relevant and selected)
- TN represents true negative (i.e. Item is not relevant and not selected)
- FP shows false positive (i.e. Item is not relevant but selected)

---

[3] Root of the Mean Square Error

- FN indicated false negative (i.e. Item is relevant but not selected)

Precision measures the accuracy of the recommendation system, which is proportionate of corrected recommended information to the total recommended information. The precision of any RS indicates how good the recommending model is in predicting positive class. It can be referred to as positive predictive values. Whereas Recall shows the recommendation system recall rate that is the ratio between the numbers of corrected information recommended to the total corrected information in the dataset. It is the same as the RS sensitivity. Accuracy (ACC) is the fraction of rightly predicted ratings in which the proposed model has predicted the right to total ratings [50].

The F-Measure criterion is obtained based on the Precision and Recall criteria, as follows [62]:

$$F1 = \frac{2 \cdot \text{Precision} \cdot \text{Recall}}{\text{Precision} + \text{Recall}} \quad (34).$$

Given the $N$ actual/predicted rating pairs ($p_{u,I}$, $r_{u,i}$), where $u$ refers to a user and $I$ to an item, the RMSE (Root Mean Squared Error) of the $N$ pairs is evaluated as [50, 63]:

$$RMSE = \sqrt{\frac{\sum_{i=1}^{N}(p_{u,i} - r_{u,i})^2}{N}} \quad (35).$$

Note that lower RMSE value indicates more accurate predictions, signifying better performance of recommender systems. Next, the efficiency of the algorithm presented in this research (UTV) undergoes a comparison in three parts. According to paper [50] In order to perform the best estimation of accuracy of the proposed model, experiments are carried out 10 times with a random 80/20 split of train/ test data. The average value of RMSE is noted. The decrease in RMSE shows the better performance of the model.

## 5-3- Comparison of UTV to Other Recommender System Models:

In this subsection, UTV is compared to well-known, standard recommender system algorithms, including the Collaborative Filtering (CF) standard [62, 64] Collaborative (CL) systems, and Content-Based Filtering (CBF).

Content-Based Filtering (CBF) calculates a degree of similarity between the users and the items to be recommended. The process is carried out by comparing the features of the item concerning the user's preferences [65]. Collaborative (CL) systems make recommendations based on groups of users with similar preferences. The similarity between users is normally computed by comparing the ratings that they give to some of the items [66]. Next, the results obtained over the MovieLens 1M and MovieLens 10M datasets are shown in Figures 5 and 6.

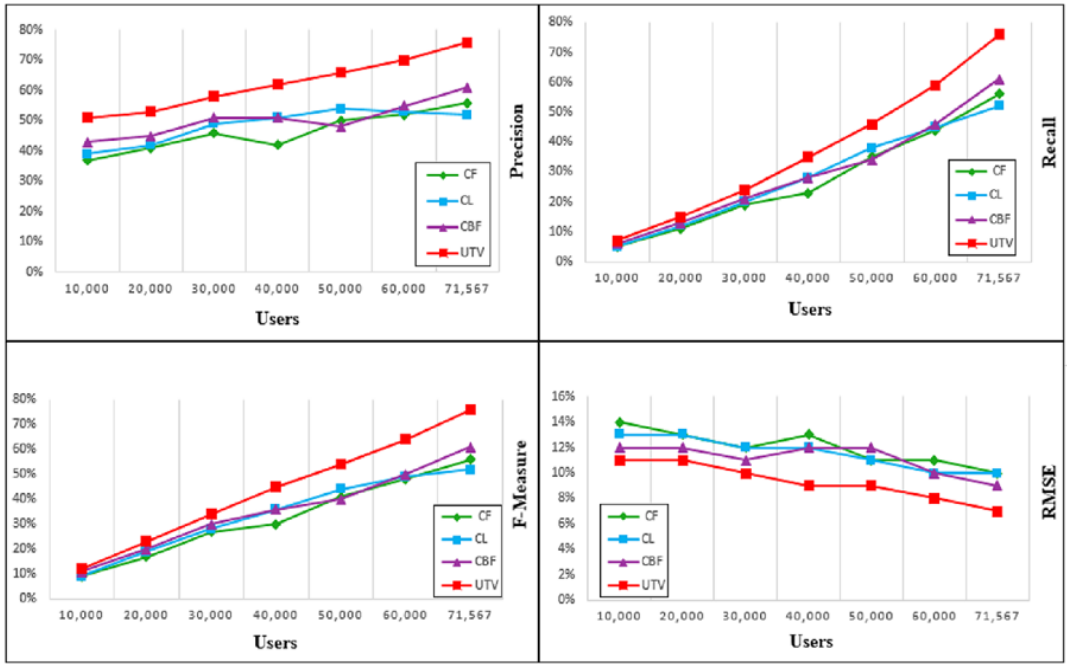

**Fig. 5. Comparison of the accuracy of the UTV algorithm to that of the others over the MovieLens 10M dataset**

The results obtained over both datasets on all the four examined criteria suggest the higher efficiency of UTV than that of the other recommender system algorithms.

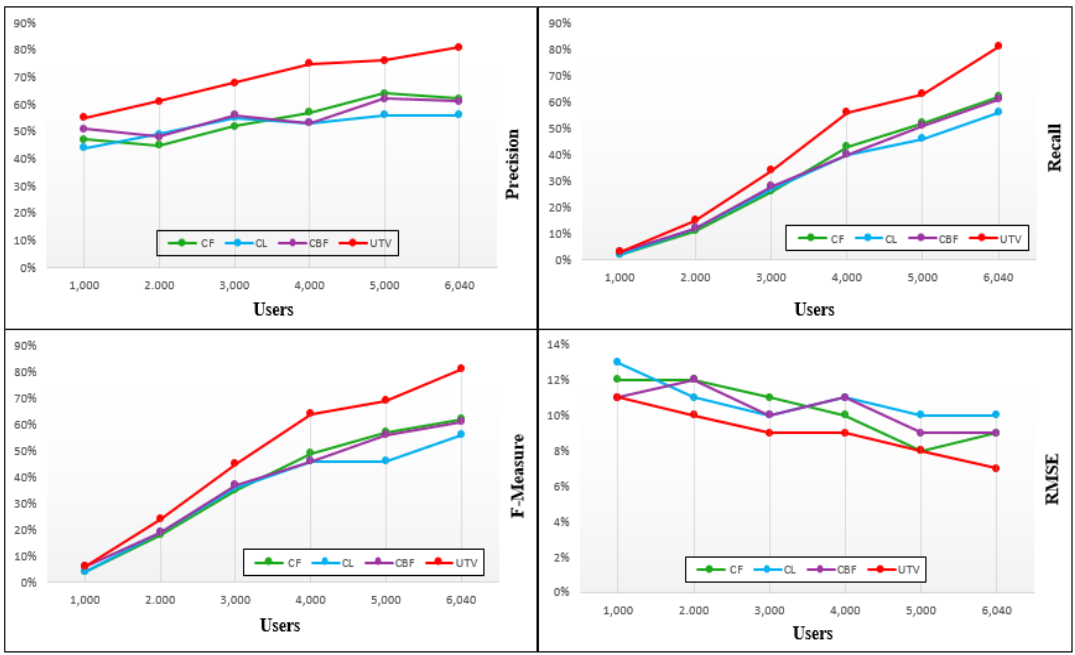

**Fig. 6. Comparison of the accuracy of the UTV algorithm to that of the others over the MovieLens 1M dataset**

## 5-4- Comparison of UTV to Other Models:

In this subsection, the method presented in this research is compared to a number of other models presented for recommender systems based on the RMSE and Accuracy criteria over MovieLens 1M and MovieLens 10M, as follows: In the paper [66], authors propose a Content-based Recommender System that exploits knowledge graph embedding for representing items. The embedding are built by leveraging on triples extracted from Wikidata and their approach for computing a user profile based on knowledge base embedding. Therefore, several approaches for building the user profiles and for generating a list of suggestions for the user are proposed. The paper [67] considers the knowledge graph as the source of side information. To address the limitations of existing embedding-based and path-based methods for the knowledge-graph-aware recommendation, researchers propose RippleNet, an end-to-end framework that naturally incorporates the knowledge graph into recommender systems. RippleNet stimulates the propagation of user preferences over the set of knowledge entities by automatically and iteratively extending a user's potential interests along with links in the knowledge graph. In the paper [68], a hybrid evolutionary-based method is presented for item clustering and production of the most suitable clusters in the offline phase of the collaborative filtering Recommender System (RS). The proposed method is a combination of the Genetic Algorithm (GA) and the Gravitational Emulation Local Search (GELS) algorithm. Next, the methods mentioned above are compared to UTV, proposed in this research. For a more accurate examination of the models, the datasets are divided into different ranges (in terms of the number of examined users) in the first step. The algorithms relevant to each model are run in the generated ranges, and the results for each range are then obtained, followed by a calculation of the mean results for each model. In the second step, the models are compared, where only the results for which the user has given the recommended item the maximum score (5 out of 5) in his rating are considered as acceptable recommendations. Therefore, the results obtained for the RMSE and Accuracy criteria over the MovieLens 1M and MovieLens 10M datasets are shown in Tables 3, 4, 5, and 6, given that the final item must have the maximum score in the user's rating.

Table 3. Comparison of UTV with the other models in terms of RMSE (**MovieLens 1M Datasets**)

| Models Based RMSE - Movielens 1M | Number of Users | | | | | | AVG |
|---|---|---|---|---|---|---|---|
| | 1,000 | 2,000 | 3,000 | 4,000 | 5,000 | 6,040 | |
| linked open data-based recommender systems[66] | 2.63 | 2.25 | 2.55 | 2 | 1.88 | 1.6 | 2.151 |
| RippleNet [67] | 2.42 | 2.44 | 2.33 | 1.9 | 2.1 | 1.46 | 2.108 |
| Genetic Algorithm and Gravitational Emulation [68] | 3.1 | 2.8 | 2.6 | 2.35 | 2.3 | 1.7 | 2.475 |
| UTV | 2.15 | 1.88 | 1.8 | 1.66 | 1.44 | 1.13 | 1.676 |

Table 4. Comparison of UTV with the other models in terms of RMSE (**MovieLens 10M Datasets**)

| Models Based RMSE - Movielens 10M | Number of Users | | | | | | | AVG |
|---|---|---|---|---|---|---|---|---|
| | 10,000 | 20,000 | 30,000 | 40,000 | 50,000 | 60,000 | 71,567 | |
| linked open data-based recommender systems | 2.22 | 2.85 | 3.39 | 2.91 | 2.35 | 2.22 | 1.9 | 2.548 |
| RippleNet | 2.1 | 2.9 | 2.66 | 2.72 | 2.54 | 2.16 | 1.77 | 2.407 |
| Genetic Algorithm and Gravitational Emulation | 2.5 | 3.2 | 3.55 | 3.15 | 2.88 | 2.77 | 2.23 | 2.897 |
| UTV | 1.8 | 2.25 | 2.11 | 1.82 | 1.62 | 1.55 | 1.44 | 1.798 |

Table 5. Comparison of UTV with the other models in terms of ACC (**MovieLens 1M Datasets**)

| Models Based ACC - Movielens 1M | Number of Users | | | | | | AVG |
|---|---|---|---|---|---|---|---|
| | 1,000 | 2,000 | 3,000 | 4,000 | 5,000 | 6,040 | |
| linked open data-based recommender systems | 0.216 | 0.277 | 0.25 | 0.288 | 0.314 | 0.32 | 0.2775 |
| RippleNet | 0.253 | 0.263 | 0.285 | 0.31 | 0.339 | 0.346 | 0.2993 |
| Genetic Algorithm and Gravitational Emulation | 0.188 | 0.218 | 0.2 | 0.225 | 0.271 | 0.292 | 0.2323 |
| UTV | 0.291 | 0.313 | 0.342 | 0.366 | 0.385 | 0.415 | 0.352 |

Table 6. Comparison of UTV with the other models in terms of ACC (**MovieLens 10M Datasets**)

| Models Based ACC - Movielens 10M | Number of Users | | | | | | | AVG |
|---|---|---|---|---|---|---|---|---|
| | 10,000 | 20,000 | 30,000 | 40,000 | 50,000 | 60,000 | 71,567 | |
| linked open data-based recommender systems | 0.177 | 0.212 | 0.261 | 0.305 | 0.366 | 0.384 | 0.421 | 0.3037 |
| RippleNet | 0.163 | 0.228 | 0.278 | 0.322 | 0.34 | 0.391 | 0.433 | 0.3078 |
| Genetic Algorithm and Gravitational Emulation | 0.138 | 0.179 | 0.231 | 0.277 | 0.329 | 0.353 | 0.386 | 0.2704 |
| UTV | 0.225 | 0.255 | 0.31 | 0.363 | 0.411 | 0.452 | 0.515 | 0.3615 |

As observed, UTV exhibits better performance than all the other models with averages of RMSE = 1.676 and ACC = 0.352 over MovieLens 1M and with averages of RMSE = 1.798 and ACC = 0.3615 over MovieLens 10M, and the RippleNet algorithm is the second best. Furthermore, UTV outperforms the other models in the point-to-point comparison of performance in all the examined ranges, which indicates the acceptable efficiency of the method presented in this research.

### 5-4-1- Top-N Recommendation:

In the next step of evaluation of the presented method, a criterion known as AUC, employed in [69], is used for its comparison to the other methods. Including rating-prediction tasks, we also test our model on top-N item-recommendation tasks. To predict a list of items, we arranged all items order by their prediction ratings and the probability of making this prediction. We evaluate UTV using the metric of the area under the ROC curve (AUC) prediction quality. AUC was defined as follows:

$$AUC = \frac{\sum ins_{i \in PositiveVecClass} rank_{ins_i} - \frac{M \times (M+1)}{2}}{M \times N} \quad (36)$$

Where M is a number of items that are rated, and N is the number of items that are not rated. Positive VecClass contains a class, and the items included in it given scores of only 4 or 5. On that basis, movies or items are acceptable for the recommendation that is scored 4 or 5 in the ratings provided by the users. A higher value of the AUC indicates better quality. The trivial AUC of a random guess method is 0:5, and the best achievable quality is 1. Results of top-N item-recommendation tasks are displayed in Tables 7 and 8. We use our model on the top-N item-recommendation task, which is designed for rating-prediction tasks. On that basis, items are predicted in this subsection that is to be recommended to the user, and the scores provided for those items are taken into account. The higher the score provided for the recommended item, the more efficient the method under examination. The results have demonstrated the effectiveness of rating embedding.

Table 7. Comparison of UTV with the other models in terms of AUC (**MovieLens 1M Datasets**)

| Models Based AUC- Movielens 1M | Number of Users | | | | | | AVG |
|---|---|---|---|---|---|---|---|
| | 1,000 | 2,000 | 3,000 | 4,000 | 5,000 | 6,040 | |
| linked open data-based recommender systems | 0.515 | 0.588 | 0.5771 | 0.6165 | 0.6547 | 0.6288 | 0.5966 |
| RippleNet | 0.6132 | 0.6433 | 0.658 | 0.643 | 0.6682 | 0.6852 | 0.6518 |
| Genetic Algorithm and Gravitational Emulation | 0.621 | 0.6175 | 0.6278 | 0.6551 | 0.6476 | 0.6781 | 0.6411 |
| UTV | 0.685 | 0.7024 | 0.7234 | 0.7461 | 0.7663 | 0.7807 | 0.7339 |

Table 8. Comparison of UTV with the other models in terms of AUC (**MovieLens 10M Datasets**)

| Models Based AUC - Movielens 10M | Number of Users | | | | | | | AVG |
|---|---|---|---|---|---|---|---|---|
| | 10,000 | 20,000 | 30,000 | 40,000 | 50,000 | 60,000 | 71,567 | |
| linked open data-based recommender systems | 0.633 | 0.6553 | 0.6632 | 0.6478 | 0.6900 | 0.7114 | 0.7053 | 0.6723 |
| RippleNet | 0.6543 | 0.668 | 0.6577 | 0.6825 | 0.7221 | 0.7122 | 0.7014 | 0.6854 |
| Genetic Algorithm and Gravitational Emulation | 0.5899 | 0.6223 | 0.6681 | 0.6721 | 0.6655 | 0.7226 | 0.7005 | 0.663 |
| UTV | 0.7044 | 0.7228 | 0.7558 | 0.7893 | 0.7940 | 0.8111 | 0.8355 | 0.7732 |

The results obtained in this subsection over the two datasets also indicate the greater efficiency of the method presented in this research than that of the other compared algorithms.

### 6- Conclusion:

In this research, performed on the MovieLens datasets, a movie recommender system to users was proposed using ensemble learning and graph embedding. The study was conducted in two general steps. In the first step, a number of classes were created using ensemble learning, and different modes of user behavior were analyzed and compared to target user behavior through a combination of the classes so that classes most similar to the target user (in terms of profile and behavior) could be detected. Then, each of the obtained combinations was represented as a fuzzy rule, and the similarity of the users to the target user in each of the fuzzy rules was obtained. Next, the obtained results were displayed as a decision tree, and three classes of users most similar to the target user were selected. In the second step, a heterogeneous knowledge graph was generated using the items liked by the target user and three classes of users most similar to him, and each of the items selected by the users was represented by a number of nodes and various links established between different nodes. Then, the fuzzy rules were extracted based on the relations defined between the nodes (the edges created between the pairs of nodes). Next, a UTV was generated for each user based on the extracted fuzzy rules, composed of a combination of vectors as (SF & SFS), (SMPF & SFS), and (SMPSF & SFS), and recommendations on the outputs of the three vectors were made to the user according to the results obtained from them. Finally, the proposed method was evaluated, and the obtained results demonstrated that the method used in this research exhibited higher accuracy and efficiency than those in previous works in their recommendations made to users. In the evaluation part, the MovieLens datasets were divided into different ranges, and the results were obtained for each range based on the examined criteria, followed by calculations of the mean result for each range. The UTV algorithm exhibited better performance than all the other compared algorithms also in this part, where the obtained results demonstrated that its overall performance was better than that of all the other algorithms, and its resulting outputs exhibit higher accuracy than in the other methods.